\documentclass{article}
\usepackage{spconf,amsmath,graphicx}
\usepackage{xurl}
\usepackage{hyperref}
\usepackage{booktabs,multirow}
\usepackage{CJKutf8}
\usepackage{amssymb}
\usepackage{orcidlink}
\newcommand{\ourmethod}{SemiCodec}
\newcommand{\demourl}{https://pennyw1-se.github.io/SemiCodec-Demo-Page/}

\title{{\Large \textsc{\ourmethod}}: Bridging Speech Tokenization and High-Fidelity Reconstruction with Semi-Discrete Representations
}
\name{%
  \parbox{\linewidth}{%
    \centering
    \itshape
    Kecan Mao$^{1\star}$~\orcidlink{0009-0007-7936-3603} \qquad 
    Nan Xu$^{2}$~\orcidlink{0009-0006-1010-0381} \qquad
    Zhaoyang Meng$^{1}$~\orcidlink{0009-0007-0417-877X} \qquad
    Zhengyao Ma$^{1}$~\orcidlink{0009-0003-3873-7745} \qquad \\[2pt]
    Yingming Gao$^{1}$~\orcidlink{0000-0001-5881-3723} \qquad
    Yi Zhao$^{2}$~\orcidlink{0000-0002-3555-9408} \qquad 
    Ya Li$^{1\dagger}$~\orcidlink{0000-0002-6284-5039} \qquad
  }%
}
\address{
  \vspace{-1em}
    \thanks{%
    \begin{tabular}[t]{@{}l@{}}
        $^{\star}$Work performed during the internship at Tencent.\\
        $^{\dagger}$Corresponding author. Email: yli01@bupt.edu.cn
    \end{tabular}} \\ 
  $^{1}$School of Artificial Intelligence, Beijing University of Posts and Telecommunications, Beijing, China \\
  $^{2}$WeChat, Tencent Inc., Beijing, China 
}
\begin{document}
\ninept
\maketitle
\begin{abstract}
Speech representation learning requires a discrete modeling interface while preserving high waveform fidelity. Existing
approaches rarely achieve both: discrete tokens ease modeling but lose
fine-grained acoustic information, while continuous latents preserve fidelity
but require specialized generative objectives. We introduce \textbf{SemiCodec},
a novel \textbf{Semi}-discrete neural speech \textbf{Codec} with synchronized
discrete tokens and continuous latents. Direct joint
optimization may create a reconstruction shortcut through the continuous
pathway, leaving the discrete tokens insufficiently grounded. To address this
issue, we propose a \textbf{Preserve-then-Ground} training paradigm that first establishes a
high-fidelity synthesis space and then grounds SSL-guided discrete tokens within it. Experiments show that
SemiCodec outperforms fully discrete codecs in reconstruction quality and
intelligibility while remaining competitive with continuous representations
in acoustic fidelity and speaker preservation. Audio samples are available at
\expandafter\url\expandafter{\demourl}.
\end{abstract}
\section{Introduction}

% Recent advances in generative AI have established tokenization as a critical
% interface between high-dimensional data and generative models. A desirable
% tokenizer should produce representations that are easy to model while
% preserving sufficient information for faithful reconstruction. This dual
% requirement is particularly challenging for speech, where linguistic content
% is intertwined with speaker identity, prosody, and fine-grained acoustic
% variation.

Speech combines discrete linguistic content with continuously varying acoustic
information~\cite{ploujnikov2026hybridcodec}: linguistic content can be
organized into discrete units, whereas speaker identity, prosody, emotion, and
acoustic details vary continuously. An effective speech representation should
therefore support both efficient modeling and faithful reconstruction.

Existing speech codecs predominantly follow discrete or continuous routes.
Discrete methods, such as DAC~\cite{kumar2023high}, Mimi~\cite{defossez2024moshi}, WavTokenizer~\cite{ji2025wavtokenizer}, and
X-Codec~\cite{ye2025xcodec}, provide compact tokens suitable for next-token
prediction. DualCodec~\cite{li2025dualcodec}, SAC~\cite{chen2025sac}, and OmniCodec ~\cite{hu2026omnicodec} further
separate semantic and acoustic token streams, but both remain quantized and may
lose acoustic detail at low frame rates. In contrast, continuous representations,
including Semantic-VAE~\cite{niu2025semanticvae}, Ming-UniAudio~\cite{yan2025minguniaudio} and dots.tts~\cite{lian2026dotstts} avoid hard quantization and better
preserve acoustic variation in continuous latent spaces. However, continuous representations typically
require regression, flow matching, diffusion, or specialized continuous
autoregressive objectives. These complementary limitations motivate a
representation that retains a compact discrete interface without
forcing fine-grained acoustic information through the same quantization
bottleneck.

\begin{figure}[t]
    \centering
    \includegraphics[width=0.48\textwidth]{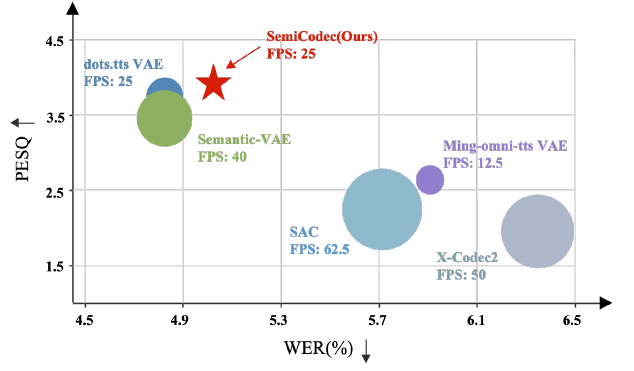}
    \caption{Comparison of codecs on speech reconstruction. The x-axis shows WER, reflecting speech intelligibility, while the y-axis presents PESQ-WB, reflecting  perceptual quality of speech reconstruction. Circle size indicates temporal frame rate.}
    \label{fig:Codec_compare}
    \vspace{-4mm}
\end{figure}
\vspace{-4mm}

Following this motivation, we propose
\textbf{SemiCodec}, a \textbf{Semi}-discrete neural speech \textbf{Codec} that
uses discrete tokens as an explicit interface and continuous latents for
preserving acoustic details. The two streams are mapped into a
shared synthesis space for waveform reconstruction. Naive joint optimization,
however, will allow the information-rich continuous pathway to become a
reconstruction shortcut, thus causing the decoder to underuse the discrete tokens.

\begin{figure*}[t!]
    \centering
    \includegraphics[width=\textwidth, height=10.5cm, keepaspectratio]{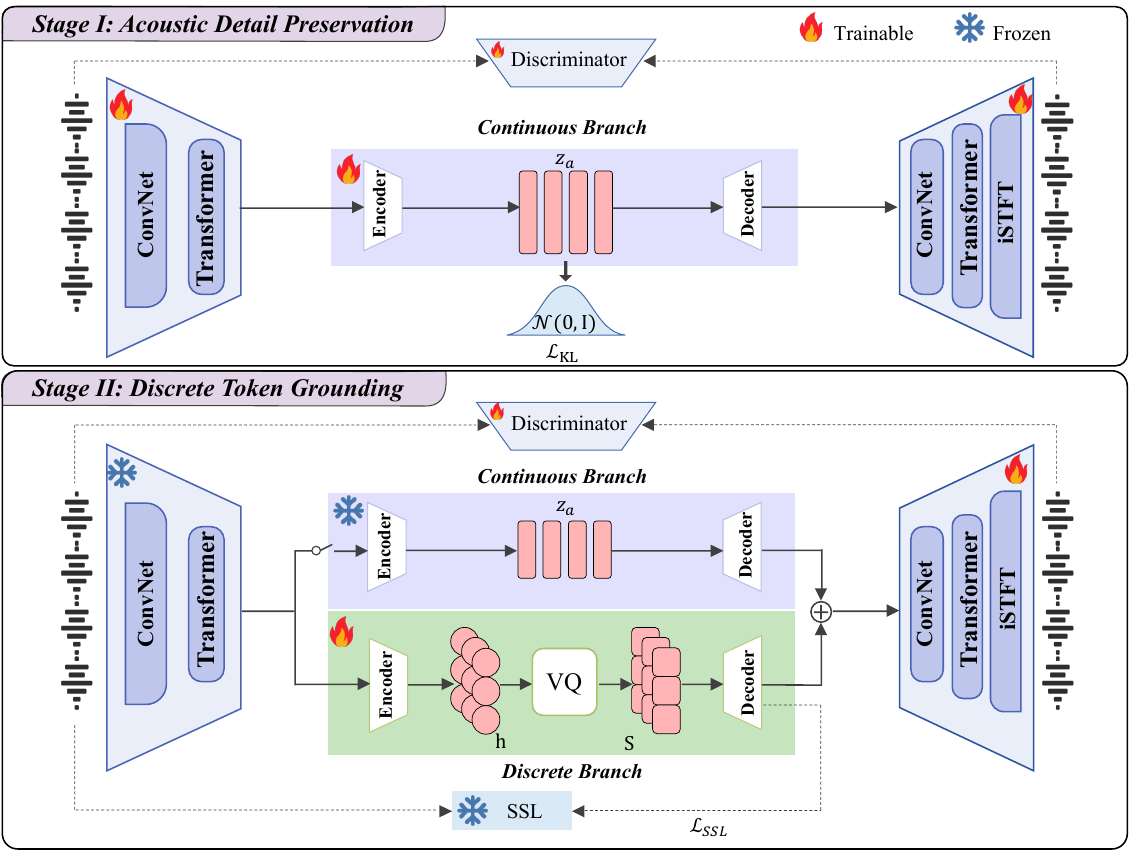}
    \caption{Overview of SemiCodec and its \textbf{Preserve-then-Ground}
    training paradigm. The discrete and continuous branches encode
    SSL-guided and fine-grained acoustic information, respectively, and the
    decoded features are added in a shared synthesis space for waveform
    reconstruction. Stage I trains the Continuous Branch; Stage II freezes
    the learned representation and trains the Discrete Branch with
    waveform and SSL supervision. The
    \protect\raisebox{-0.15em}{\includegraphics[height=0.9em]{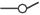}}
    symbol denotes Acoustic Dropout, which encourages the Waveform Decoder to
    use the discrete stream.}
    \label{fig:arch}
\end{figure*}

To address this challenge, we introduce a \textbf{Preserve-then-Ground} training paradigm. We first train the
continuous pathway to establish a high-fidelity synthesis space, then freeze
the learned acoustic representation and ground the discrete tokens in that
space under self-supervised guidance. This staged optimization
mitigates the continuous-path shortcut while retaining a compact discrete
interface and high-quality reconstruction.

Our main contributions are summarized as follows:
\begin{itemize}
    \item We propose \textbf{SemiCodec}, a semi-discrete speech codec that
    combines SSL-guided discrete tokens with continuous latents for high-fidelity reconstruction.
    \item We identify the continuous-path reconstruction shortcut in joint
    training and introduce \textbf{Preserve-then-Ground}, a staged paradigm
    that effectively grounds the discrete tokens without degrading the
    pretrained acoustic representation.
    \item SemiCodec \textbf{outperforms fully discrete codecs} in
    reconstruction quality while remaining competitive with
    continuous representations (see Figure~\ref{fig:Codec_compare}). Ablation and codebook-usage results further
    validate the proposed training paradigm.
\end{itemize}

\section{Method: SemiCodec}
\label{sec:method}

\subsection{Overview}
\label{ssec:overview}

As shown in Figure~\ref{fig:arch}, SemiCodec encodes a waveform $\mathbf{x}$ into
synchronized SSL-guided discrete tokens $\mathbf{s}$ and continuous acoustic
latents $\mathbf{z}_a$. The two streams are projected into a shared synthesis
space, fused by addition, and decoded as
\begin{equation}
    \hat{\mathbf{x}}
    =G\!\left(D_s(\mathbf{s})+D_a(\mathbf{z}_a)\right),
    \label{eq:reconstruction}
\end{equation}
where $D_s$, $D_a$, and $G$ denote the Discrete-Branch, Continuous-Branch, and
waveform decoders, respectively. To prevent the continuous path from bypassing
the discrete tokens, our \textbf{Preserve-then-Ground} paradigm first preserves acoustic
details and then grounds the discrete tokens in the learned synthesis space.

\subsection{Waveform Encoder and Decoder}
\label{ssec:waveform_codec}

The shared encoder applies causal strided convolutions followed by a causal
Qwen2-style Transformer backbone~\cite{yang2024qwen2}, following
TS3-Codec~\cite{wu2025ts3codec}, and then produces $\mathbf{h}$ at 25 frames per
second. The waveform decoder first upsamples the fused representation with a transposed
convolution, followed by a Qwen2-style Transformer backbone.
Following Vocos~\cite{siuzdak2024vocos}, an iSTFT head predicts the complex
spectrum and reconstructs the waveform directly, without an intermediate
mel-spectrogram or external vocoder.

\subsection{Semi-discrete Representation}
\label{ssec:semi_discrete_representation}

\noindent\textbf{Continuous Branch.}
We use a variational bottleneck~\cite{kingma2014auto}. The continuous encoder
$E_a(\cdot)$ predicts a diagonal Gaussian posterior and samples a
64-dimensional latent through reparameterization:
\begin{equation}
\begin{split}
    [\boldsymbol{\mu},\log\boldsymbol{\sigma}^{2}]&=E_a(\mathbf{h}),\\
    \mathbf{z}_{a}&=\boldsymbol{\mu}
    +\boldsymbol{\sigma}\odot\boldsymbol{\epsilon},\quad
    \boldsymbol{\epsilon}\sim\mathcal{N}(\mathbf{0},\mathbf{I}).
\end{split}
\end{equation}
Here, $\boldsymbol{\mu}$ and $\boldsymbol{\sigma}$ are the posterior mean and
standard deviation, $\boldsymbol{\epsilon}$ is standard Gaussian noise, and
$\odot$ denotes element-wise multiplication. The decoder $D_a(\cdot)$ projects
$\mathbf{z}_a$ into the shared synthesis space.

\noindent\textbf{Discrete Branch.}
A frame-wise encoder and a vector quantizer
~\cite{oord2017vqvae} map $\mathbf{h}$ to one token per frame. The quantizer
projects each input feature $\mathbf{h}_t$ into a
lookup feature $\mathbf{e}_t$ and uses a codebook of $K=4{,}096$ entries. After
L2 normalization, the token index is selected by
\begin{equation}
    s_t=\arg\min_{k\in\{1,\ldots,K\}}
    \left\|\bar{\mathbf{e}}_t-\bar{\mathbf{c}}_k\right\|_2^2,
    \label{eq:factorized_vq}
\end{equation}
where $t$ is the frame index, $k$ indexes the codebook, $\mathbf{c}_k$ is its
$k$-th entry. The selected embedding
is then projected back to the model dimension and mapped by $D_s(\cdot)$ into the
shared synthesis space. A straight-through estimator is used to propagate gradients
through the discrete assignment.

The two branches provide complementary interfaces at the same 25-Hz frame
rate. The Discrete Branch represents each frame with one of 4,096 symbols,
forming a compact discrete interface, whereas the
64-dimensional continuous latent retains information that is difficult to
express through a single codebook entry, such as speaker characteristics,
prosody, and local acoustic variation. Their synchronization permits direct
frame-wise fusion. Mapping both streams to the same synthesis space and adding
them also keeps the input dimension of the Waveform Decoder fixed, while
allowing either stream to contribute information at every frame.

Direct joint training creates an asymmetric optimization problem. Because the
continuous latent is not constrained to select among a finite set of entries,
it can reduce the reconstruction loss more rapidly during early training. The
Waveform Decoder may consequently learn to depend on this path, leaving the
Discrete Branch with weak reconstruction feedback. SSL distillation alone
does not fully resolve this issue: it can align discrete features with the SSL
teacher without ensuring that those features are used for waveform synthesis.
Our \textbf{Preserve-then-Ground} paradigm separates these objectives. It first
constructs a stable, high-fidelity acoustic synthesis space and subsequently
trains the discrete tokens to enter and complement that space.
\subsection{\textbf{Preserve-then-Ground} Training}
\label{ssec:training_strategy}

\subsubsection{Training Objectives}
\label{sssec:training_objective}
Following DAC~\cite{kumar2023high} and HiFi-GAN~\cite{kong2020hifi}, we define
the waveform objective as
\begin{equation}
    \mathcal{L}_{\mathrm{wav}}
    =\lambda_{\mathrm{mel}}\mathcal{L}_{\mathrm{mel}}
    +\lambda_{\mathrm{adv}}\mathcal{L}_{\mathrm{adv}}
    +\lambda_{\mathrm{feat}}\mathcal{L}_{\mathrm{feat}},
    \label{eq:waveform_objective}
\end{equation}
where $\mathcal{L}_{\mathrm{mel}}$ is a multi-scale mel-spectrogram L1 loss,
$\mathcal{L}_{\mathrm{adv}}$ is the generator adversarial loss, and
$\mathcal{L}_{\mathrm{feat}}$ is the L1 feature-matching loss. We use a least-squares GAN objective with a
Multi-Period Discriminator (MPD) and a Multi-Resolution complex-spectrogram
Discriminator (MRD).

The continuous posterior is regularized toward a standard Gaussian prior:
\begin{equation}
    \mathcal{L}_{\mathrm{KL}}
    =D_{\mathrm{KL}}\!\left(q_a(\mathbf{z}_a\mid\mathbf{h})
    \,\|\,\mathcal{N}(\mathbf{0},\mathbf{I})\right),
    \label{eq:kl_loss}
\end{equation}
where $q_a(\mathbf{z}_a\mid\mathbf{h})$ is the posterior predicted by
$E_a(\cdot)$ and $D_{\mathrm{KL}}$ denotes Kullback--Leibler divergence.

For the discrete representation, following RepCodec
~\cite{huang2024repcodec} and DualCodec~\cite{li2025dualcodec}, we distill
features from a frozen w2v-BERT 2.0 encoder~\cite{barrault2023seamless}. Its
17th layer's output is temporally aligned with the decoded discrete representation
$\mathbf{h}_s=D_s(\mathbf{s})$, yielding
\begin{equation}
    \mathcal{L}_{\mathrm{SSL}}
    =\frac{1}{T}\sum_{t=1}^{T}
    \left(1-\cos\!\left(\mathbf{h}_{s,t},
    \widetilde{\mathbf{y}}_t\right)\right),
    \label{eq:ssl_distillation}
\end{equation}
where $T$ is the sequence length, $\mathbf{h}_{s,t}$ is the discrete feature at
frame $t$, $\widetilde{\mathbf{y}}_t$ is the aligned teacher feature. The quantizer is optimized with
\begin{equation}
    \mathcal{L}_{\mathrm{VQ}}
    =\lambda_{\mathrm{com}}\mathcal{L}_{\mathrm{com}}
    +\lambda_{\mathrm{cb}}\mathcal{L}_{\mathrm{cb}},
    \label{eq:vq_objective}
\end{equation}
where $\mathcal{L}_{\mathrm{com}}$ and $\mathcal{L}_{\mathrm{cb}}$ are the
commitment and codebook losses.

\begin{table*}[!t]
\centering
\small
\caption{Speech reconstruction results on LibriSpeech \textit{test-other}. FPS denotes
representation frame rate; ``--'' denotes an unreported metric.}
\label{tab:reconstruction}
\vspace{1mm}
\begin{tabular*}{\textwidth}{@{\extracolsep{\fill}}l c c cc c c c c c@{}}
\toprule
\multirow{2}{*}{\textbf{Model}}
  & \multirow{2}{*}{\textbf{SR}}
  & \multirow{2}{*}{\textbf{FPS}}
  & \multicolumn{2}{c}{\textbf{PESQ$\uparrow$}}
  & \multirow{2}{*}{\textbf{STOI$\uparrow$}}
  & \multirow{2}{*}{\textbf{ViSQOL$\uparrow$}}
  & \multirow{2}{*}{\textbf{SIM$\uparrow$}}
  & \multirow{2}{*}{\textbf{WER (\%)$\downarrow$}}
  & \multirow{2}{*}{\textbf{MUSHRA$\uparrow$}} \\
\cmidrule(lr){4-5}
 & & & NB & WB & & & & & \\
\midrule
Ground Truth     & --     & --           & 4.55 & 4.64 & 1.000 & 5.00 & 1.000 & 4.66 & 90.27±2.03   \\
\midrule
\multicolumn{10}{l}{\textit{Discrete tokens}} \\
% XY-Tokenizer     & 16~k   & 100          & 2.80 & 2.23 & 0.890 & 3.46 & 0.820 & 6.19 &   \\
WavTokenizer     & 16~k   & 75           & 2.40 & 1.96 & 0.870 & 3.08 & 0.68 & 12.28 & 70.88±1.09 \\
X-codec2          & 16~k   & 50           & 2.83 & 2.26 & 0.900 & 3.05 & 0.81 & 6.46 & 71.14±2.37  \\
SAC              & 16~k   & 62.5         & 2.92 & 2.39 & 0.900 & 3.25 & 0.85 & 5.71 & 74.39±1.72   \\
\midrule
\multicolumn{10}{l}{\textit{Continuous representations}} \\
Semantic-VAE      & 16~k   & 40           & 3.99 & 3.80 & 0.969 & \underline{3.84} & \underline{0.96} & \textbf{4.81} & 77.24±2.24 \\
% MingTok-Audio    & 16~k   & 50           & \textbf{4.23} & \textbf{4.12} & \textbf{0.981} & 3.75 & 0.950 & 4.27 &   \\
Ming-omni-tts VAE        & 44.1~k   & 12.5 & 3.28 & 2.64   & 0.917 & 3.21 &  0.84   & 5.97 & 77.28±2.50  \\
dots.tts VAE & 48~k & 25 & 4.09 & 3.95 & \underline{0.973} & \textbf{3.98} & \textbf{0.97} & \textbf{4.81} &  82.83±1.65 \\
\midrule
\multicolumn{10}{l}{\textit{Semi-discrete representations}} \\
SemiCodec          & 24~k   & 25           & \textbf{4.15}      & \textbf{4.03}     & \textbf{0.975}      & 3.77    & \underline{0.96}       &  \underline{5.02}    & \textbf{86.95±1.35}\\
\bottomrule
\end{tabular*}
\vspace{-4mm}
\end{table*}

\begin{table}[!t]
    \centering
    \small
    \renewcommand{\arraystretch}{1.2}
    \caption{Ablation study of the architecture and training paradigm.}
    \label{tab:ablation}
    \vspace{1mm}
    \begin{tabular*}{\columnwidth}{@{\extracolsep{\fill}}lcccc@{}}
        \toprule
        \multirow{2}{*}{\textbf{Variant}}
        & \multicolumn{2}{c}{\textbf{PESQ$\uparrow$}}
        & \multirow{2}{*}{\textbf{ViSQOL$\uparrow$}}
        & \multirow{2}{*}{\textbf{WER(\%)$\downarrow$}} \\
        \cmidrule(lr){2-3}
        & NB & WB & & \\
        \midrule
        Joint Training & 3.87 & 3.58 & 3.58 & 5.15 \\
        Stage I & 4.12 & 3.99 & 3.73 & 5.09 \\
        Stage II & 4.14 & 4.02 & 3.76 & 5.03 \\
        \quad + Acoustic Dropout & \textbf{4.15} & \textbf{4.03} & \textbf{3.77} & \textbf{5.02} \\
        \bottomrule
    \end{tabular*}
    \vspace{-4mm}
\end{table}

\subsubsection{Stage I: Acoustic Detail Preservation}
In Stage I, we disable the Discrete Branch and optimize the waveform encoder,
Continuous Branch, and waveform decoder using
\begin{equation}
    \mathcal{L}_{\mathrm{S1}}
    =\mathcal{L}_{\mathrm{wav}}
    +\lambda_{\mathrm{KL}}\mathcal{L}_{\mathrm{KL}}.
\end{equation}
Without competition from a randomly initialized discrete path, this stage first
learns a stable synthesis space that preserves fine-grained acoustic information
and supports high-fidelity reconstruction.

\subsubsection{Stage II: Discrete Token Grounding}
Stage II initializes from Stage I and freezes the waveform encoder and
Continuous Branch, retaining them as a stable acoustic anchor. We introduce the
Discrete Branch and jointly optimize it with the waveform decoder using
\begin{equation}
    \mathcal{L}_{\mathrm{S2}}
    =\mathcal{L}_{\mathrm{wav}}
    +\lambda_{\mathrm{SSL}}\mathcal{L}_{\mathrm{SSL}}
    +\mathcal{L}_{\mathrm{VQ}}.
\end{equation}
The SSL distillation objective aligns the discrete features with the frozen
speech-teacher representation, while waveform reconstruction grounds them in the
pretrained synthesis space without disrupting the learned continuous representation.

\noindent\textit{Acoustic Dropout.}
To prevent the waveform decoder from continuing to favor the continuous path, we randomly
drop the Continuous Branch during Stage II. For each minibatch,
\begin{equation}
    \mathbf{h}_{\mathrm{dec}}=D_s(\mathbf{s})
    +mD_a(\mathbf{z}_{a}), \qquad
    m\sim\operatorname{Bernoulli}(1-p),
    \label{eq:acoustic_detail_dropout}
\end{equation}
where $p$ is the dropout probability. When $m=0$, reconstruction
must rely on the discrete stream, encouraging the decoder to use the grounded
discrete tokens.

\section{Experiments}
\label{sec:experiments}

\subsection{Experimental Setup}
\label{ssec:experimental_setup}
\textbf{Datasets.}
We train SemiCodec on approximately 795,000 hours of data, which comprises 95K hours of Emilia~\cite{he2024emilia} and LibriTTS~\cite{zen2019libritts} and 700K hours of in-house data. Audio from all sources is resampled to 24~kHz. During training, both stages use
speech segments that are randomly cropped to 9.6 seconds. This broad training
mixture is used to expose both branches to diverse linguistic and acoustic
conditions, while the fixed crop duration provides a consistent optimization
unit across the two training stages. 
% TODO: Add citations for Emilia, LibriTTS, and LibriSpeech.
% TODO: Add the duration and language composition of the internal corpus and
% the complete training mixture when the final statistics are available.

\noindent\textbf{Baselines.}
We compare SemiCodec with representative discrete and continuous speech
representations. The discrete baselines include WavTokenizer~\cite{ji2025wavtokenizer}, X-Codec2~\cite{ye2025llasa}, and SAC~\cite{chen2025sac}.
The continuous baselines include Semantic-VAE~\cite{niu2025semanticvae} and the audio VAEs adopted by
Ming-omni-tts~\cite{2026mingomnitts} and dots.tts~\cite{lian2026dotstts}. We use the released checkpoints and evaluate reconstructed waveforms with the same protocol
whenever model outputs are available.
% TODO: Add or verify citations for all baseline systems in the final draft.

\noindent\textbf{Implementation Details.}
The Discrete Branch uses a 4,096-entry codebook with 8-dimensional embeddings.
We use AdamW ($\beta=(0.8,0.9)$) with a warmup--hold--cosine schedule. The
learning rates for the main modules and discriminator are $2\times10^{-4}$ and
$1\times10^{-4}$ in Stage I; in Stage II, they are $2\times10^{-4}$,
$2\times10^{-5}$, and $4\times10^{-5}$ for the Discrete Branch, waveform
decoder, and discriminator, respectively. Training uses BF16 on 32 NVIDIA H20 GPUs. For baseline outputs available
to us, we apply the same metric implementations on the same test utterances.
% Released baselines retain their native sampling rates, and reconstructed
% waveforms are converted only when required by an evaluation metric.
% TODO: Add datasets, data statistics, baseline checkpoint
% information, evaluation protocols, and implementation details.

\subsection{Reconstruction Results}
\label{ssec:reconstruction_results}

We evaluate objective
reconstruction quality on the full LibriSpeech \textit{test-other}~\cite{panayotov2015librispeech} set, whose
challenging noisy acoustic conditions provide a robust test of content and speaker
preservation. We report narrow-band and wide-band
PESQ~\cite{rix2001perceptual}\footnote{\url{https://github.com/ludlows/python-pesq}}, 
STOI~\cite{taal2010stoi}\footnote{\url{https://github.com/mpariente/pystoi}}, 
ViSQOL~\cite{chinen2020visqol}\footnote{\url{https://github.com/google/visqol}}, 
speaker similarity (SIM), and word error rate (WER). Specifically, SIM is
computed as the cosine similarity between reference and reconstructed speech embeddings
extracted by the WavLM-Large fine-tuned speaker-verification model
(\texttt{wavlm\_large\_finetune})~\cite{chen2022wavlm}. WER is computed from
transcriptions generated by Whisper-large-v3~\cite{radford2023whisper}. We additionally conduct
a MUSHRA~\cite{schoeffler2018webmushra} listening test with 20 participants and
8 randomly selected LibriSpeech \textit{test-other}~\cite{panayotov2015librispeech} utterances.

% \vspace{-1.0em}
\noindent\textbf{Results.}
As shown in Table~\ref{tab:reconstruction}, SemiCodec achieves strong
reconstruction quality at only 25 FPS. Compared with
SAC, the strongest discrete baseline in PESQ-WB, it improves PESQ-WB from 2.39
to 4.03 and reduces WER from 5.71\% to 5.02\%, despite reducing the frame rate
from 62.5 to 25 FPS. The same trend appears in subjective evaluation: its
MUSHRA score increases from 74.39 to 86.95. These results indicate that
preserving fine acoustic information outside the quantization bottleneck can
substantially improve fidelity without increasing the discrete token rate.

At the same 25-FPS rate as dots.tts VAE, SemiCodec obtains higher PESQ-NB,
PESQ-WB, and STOI, while dots.tts retains modest advantages in ViSQOL and SIM.
SemiCodec also approaches the continuous baselines in WER and substantially
narrows the subjective-quality gap to the ground truth. Its main benefit is
therefore not uniform dominance over continuous representations, but a more
favorable interface-quality trade-off: it provides an explicit discrete stream
alongside continuous latents while retaining reconstruction quality comparable to
continuous VAEs. 
% Overall, SemiCodec provides a favorable
% balance between representation rate, acoustic fidelity, and content
% preservation, as illustrated in Fig.~\ref{fig:Codec_compare}.

\subsection{Ablation Study}
\label{ssec:ablation}

\begin{table}[t]
    \centering
    \small
    \setlength{\tabcolsep}{7pt}
    \renewcommand{\arraystretch}{1.2}
    \caption{Codebook usage on LibriSpeech \textit{test-clean}.}
    \label{tab:codebook_usage}
    \vspace{1mm}
    \begin{tabular}{lc}
        \toprule
        \textbf{Training Paradigm} & \textbf{Usage (\%)}$\uparrow$ \\
        \midrule
        Joint Training & 82.71 \\
        \textbf{Preserve-then-Ground} & \textbf{97.31} \\
         \quad w/o Acoustic Dropout & 91.67 \\
        \bottomrule
    \end{tabular}
    \vspace{-4mm}
\end{table}

Table~\ref{tab:ablation} validates each component of our training paradigm.
Compared with Joint Training, Stage I improves PESQ-WB from 3.58 to 3.99. This
large gain supports establishing the acoustic synthesis space before the two
branches compete for reconstruction. Introducing the grounded discrete tokens
in Stage II further improves PESQ-WB to 4.02, showing that the discrete stream
can complement rather than disrupt the preserved continuous representation.
Acoustic Dropout produces a smaller but consistent improvement across all four
metrics and gives the best overall result. This progression supports both the
stage ordering and the explicit pressure placed on the Waveform Decoder to use
the discrete representation.

\noindent\textbf{Codebook Usage.}
As shown in Table~\ref{tab:codebook_usage}, on LibriSpeech
\textit{test-clean}~\cite{panayotov2015librispeech},
\textbf{Preserve-then-Ground} increases codebook usage from 82.71\% to 97.31\%.
Removing Acoustic Dropout reduces it to 91.67\%, consistent with the decoder
placing less demand on the discrete pathway when the continuous features are
always available. Codebook usage provides complementary evidence that the proposed
training paradigm avoids severe under-utilization of the discrete
representation.

\vspace{-3.5mm}

\section{Conclusion}
\vspace{-3mm}
\label{sec:conclusion}

In this paper, we presented \textbf{SemiCodec}, a \textbf{Semi}-discrete neural speech
\textbf{Codec} that combines an SSL-guided Discrete Branch with a Continuous
Branch for fine-grained acoustic information. To prevent the information-rich
continuous representation from becoming a reconstruction shortcut, we
introduced the \textbf{Preserve-then-Ground} training paradigm, which first
establishes a stable acoustic synthesis space and then grounds discrete tokens
within that space. Experiments show that SemiCodec achieves superior
reconstruction quality and intelligibility over fully discrete codecs while
retaining a compact token rate, and remains competitive with continuous
representations in acoustic fidelity and speaker preservation. By providing
synchronized discrete and continuous representations within a unified codec,
SemiCodec offers a practical pathway toward joint representation modeling for
speech generation.

% Please follow the IEEE Citation Guidelines, \url{https://ieee-dataport.org/sites/default/files/analysis/27/IEEE\%20Citation\%20Guidelines.pdf} for formatting of references.


\bibliographystyle{IEEEbib}
\bibliography{strings,refs}

\end{document}